\documentclass[a4paper]{article} 
\usepackage{amsmath,amsthm,amssymb}

\title{Quantization of classical integrable systems\\
Part II:\\
quantization of functions on Poisson manifolds}
\author{M. Marino and N. N. Nekhoroshev \\
{\small Dipartimento di Matematica, Universit\`{a} degli Studi di Milano,} \\
{\small via Saldini 50, I-20133 Milano (Italy)}}

\newtheorem{thm}{Theorem}
\newtheorem{cor}[thm]{Corollary}
\newtheorem{lem}[thm]{Lemma}
\newtheorem{prop}[thm]{Proposition}

\theoremstyle{definition}
\newtheorem{defn}{Definition}[section]

\theoremstyle{remark}
\newtheorem{rem}{Remark}[section]

\def\beq{\begin{equation}}
\def\eeq{\end{equation}}

\def\lop{{\cal L}}
\def\mop{{\cal M}}
\def\aop{{\cal A}}
\def\bop{{\cal B}}
\def\cop{{\cal C}}
\def\dop{{\cal D}}
\def\nop{{\cal N}}
\def\hop{{\cal H}}
\def\sop{{\cal S}}

\def\fop{{\cal F}}
\def\rn{\mathbb{R}}
\def\nn{\mathbb{N}}
\def\zn{\mathbb{Z}}

\def\galg{\mathfrak{g}}

\def\aalg{\mathfrak{A}}

\def\ttop{{\cal T}}

\def\ipn2{\left[\frac n2\right]}

\def\Sym{{\rm Sym}}

\numberwithin{equation}{section} \numberwithin{thm}{section}

\begin{document}

\maketitle

\begin{abstract}
In a previous work we have introduced the concept of
quasi-integrable quantum system. In the present one we determine
sufficient conditions under which, given an integrable classical
system, it is possible to construct a quasi-integrable quantum
system by means of a quantization procedure based on the
symmetrized product of operators. This procedure will be applied
to concrete classes of integrable systems in two following papers.
\end{abstract}

\section{Introduction}

In a previous paper \cite{part1} (see also references therein) we
have introduced the concept of quasi-integrable quantum system. It
represents a quantum equivalent of the concept of classical
integrable system in the noncommutative sense, i.e.\ with a number
of functionally independent first integrals generally greater than
the dimension of the configuration space
\cite{TMMO,Arnold,fomenko,fasso}. As an application of this
concept, in this paper we discuss about the mathematical basis of
the quantization by symmetrization. The simplest situation, among
those considered in the paper, is the following. There exists a
one-to-one correspondence between a given set of functions on an
abstract Poisson manifold, and a given set of operators on some
configuration space. This correspondence preserves the linear
operations, and to the Poisson bracket of two functions it
associates the commutator of the corresponding operators. We then
consider the extension of this correspondence, which associates
with the (commutative) product of functions the symmetrized
product of the corresponding operators, where the product of
operators is defined as their composition. We investigate under
which conditions this extension preserves the correspondence
between the Poisson brackets of functions on one side, and the
commutators of the corresponding operators on the other. In this
way, we find conditions on a classical integrable system which
allow a corresponding quantum integrable system to be constructed
by means of symmetrization. The main condition is the existence of
a set of functions $B=(B_1,\ldots,B_l)$ on a symplectic manifold
$M^{2n}$, which is a basis of a finite dimensional Lie algebra
with respect to Poisson brackets, and moreover has the following
property. There exists an ``integrable'' set of functions of the
type considered in \cite{part1}, $F=(F_1,\ldots,F_{2n-k})$, which
are polynomials of the functions of the set $B$ with constant
coefficients: $F_j=P_j(B)$, $j=1,\ldots,2n-k$. Two main cases are
considered: the general one and the case in which
$\{B_i,B_j\}=c_{ij}$, where $c_{ij}$ are constants. In the general
case a sufficient condition for the construction of an integrable
quantum system is that the polynomials $P_1, \ldots, P_k$ are of
first degree, i.e., $F_1,\ldots,F_k$ are linearly dependent on
$B$:
\[
F_j= \sum_{i=1}^l d_{ji} B_i+ b_j \ ,
\]
where $c_{ji}$ and $b_j$ are constants. In the latter case
instead, a quadratic dependence on the functions of the set $B$ is
sufficient, that is $\deg P_i\leq 2$, $i=1,\ldots, k$. A typical
example of the second case is when $B$ is the set (or a subset) of
the linear canonical coordinates in the symplectic manifold
$\mathbb{R}^{2n}$: $B=(p,x)$. In this case $l=2n$. Note that we
fix the Planck constant $\hbar=1$ throughout the article, and
correspondingly we never consider the limit $\hbar \to 0$. In
particular, we consider only exact quantization, which means that
if three functions $A,B,C$ satisfy the equality $\{A,B\}=C$, then
their quantizations $\hat A, \hat B, \hat C$ satisfy the exact
equality $[\hat A,\hat B]=\hat C$ without any additional term
which tends to 0 as $\hbar \to 0$ (as for example in \cite{QG,
Faddeev}).

In quantum physics, the main source of integrable sets of
operators is the quantization of integrable sets of functions on
symplectic manifolds. The following statement is generally true:
if a classical system is integrable, then the corresponding
quantum system is also integrable. This refers to systems which
describe real phenomena of nature, because artificial examples of
systems which contradict this rule can be apparently constructed.
Sometimes this rule has a formal basis. In this paper we consider
some significant cases in which, starting from integrable
classical systems, one can easily construct integrable quantum
systems. A fundamental role in this construction is played by the
so-called ``symmetrization'' of products of operators
corresponding to classical functions on a symplectic manifold.
Hence we begin with the study of some general properties of
symmetrized products in an abstract associative algebra.

\section{Symmetrized products in an associative algebra}

Let us consider an arbitrary associative algebra $\mathfrak{A}$,
for example an algebra of linear operators with composition as
product. The operation of commutation $[\mathcal{A}, \mathcal{B}]:
=\mathcal{A}\mathcal{B} -\mathcal{B} \mathcal{A}$ between elements
$\mathcal{A}$ and $\mathcal{B}$ of the algebra $\mathfrak{A}$
transforms this algebra into a Lie algebra. A commutator can be
considered as (twice) the ``antisymmetrized product'' of two
elements of the algebra. It is then natural to introduce also a
``symmetrized product''.
\begin{defn}
The operation which associates with the pair $(\aop, \bop)$ the
element
\begin{equation}\label{symp}
\Sym_2(\aop, \bop):= \frac 12 (\aop \bop +\bop \aop)
\end{equation}
is called {\it symmetrized product} of the associative algebra
$\mathfrak{A}$. We shall denote this operation between two
elements also with the symbol $\diamond$:
\[
\aop \diamond\bop := \Sym_2(\aop, \bop)= \bop \diamond\aop\,.
\]
We also call {\it symmetrized product of the set} $(\aop_1,
\aop_2, \dots, \aop_k)$ of $k$ elements of the algebra
$\mathfrak{A}$ the expression
\[
\Sym_k(\aop_{1}, \aop_{2}, \dots, \aop_{k}):= \frac 1 {k!}
\sum_{\pi \in \Pi_k} {\cal A}_{\pi(1)} {\cal A}_{\pi(2)} \cdots
{\cal A}_{\pi(k)} \ ,
\]
where $\Pi_k$ is the set of all $k!$ permutations $\pi$ of $k$
objects.
\end{defn}
The symmetrized product is distributive with respect to the sum,
but it is not associative. In general we have in fact
\[
\aop_1\diamond (\aop_2\diamond \aop_3) \neq (\aop_1\diamond
\aop_2)\diamond \aop_3 \,.
\]
Moreover, both members of the above inequality are in general
different from $\Sym_3(\aop_1, \aop_2, \aop_3)$, see lemma
\ref{lem1}.

Let us consider the linear space $\sop_C^{l}= \sop_C^{0,l}$ of all
commutative polynomials $P=P(B)$ of $l$ variables. By making use
of the symmetrized product, it is possible to associate with any
element $P\in\sop_C^{0,l}$ a noncommutative polynomial $P^{\rm
sym}\in \sop_N^{l}= \sop_N^{0,l}$ (see definition 
of noncommutative polynomial in \cite{part1}) which is symmetrical
with respect to any permutation of its $l$ arguments. Consider the
linear map sym: $\sop_C^{l}\to \sop_N^{l}$, which associates with
an arbitrary monomial $Q=B_{i_1} B_{i_2} \cdots B_{i_k}$ the
symmetrized product
\[
Q^{\rm sym}:= \Sym_k(B_{i_1}, B_{i_2}, \dots, B_{i_k})\ .
\]
The set of all monomials $Q$ is a basis of the linear space
$\sop_C^{l}$. Therefore the mapping $Q\mapsto Q^{\rm sym}$ defines
a linear map sym: $P\mapsto P^{\rm sym}$ on the full space
$\sop_C^{l}$.
\begin{defn} \label{symm0}
We call the noncommutative polynomial $P^{\rm sym} \in
\sop_N^{l}$, associated with the commutative polynomial $P(B)\in
\sop_C^{l}$ by the linear map defined above, the {\it
symmetrization} of the polynomial $P=P(B_1, \ldots, B_l)$.
\end{defn}
The abelianization of $P^{\rm sym}$ (see definition 
of abelianization in \cite{part1}) obviously coincides with $P$.
The map sym defined above is a one-to-one linear map from
$\sop_C^{l}$ to $\sop_{\rm sym}^{l}$, where $\sop_{\rm sym}^{l}$
denotes the linear space of all symmetrical noncommutative
polynomials of $l$ variables. The restriction to $\sop_{\rm
sym}^{l}$ of the abelianization $T$ represents the inverse map of
sym.

The symmetrization of polynomials will play an essential role in
the following of this paper, since it is the main tool which we
shall employ for the quantization of classical hamiltonian
systems. However, before putting this tool at work, we need
preliminarily some important results about the algebraic relations
between commutators and symmetrized products in the associative
algebra $\mathfrak{A}$. In the following propositions, $\aop$ and
$\bop$ (with suitable indexes when necessary) will denote generic
elements of $\mathfrak{A}$.
\begin{prop}\label{pc1}
We have
\begin{align}
&[\aop,\Sym_k(\bop_1, \bop_2, \dots, \bop_k)]= \Sym_k([\aop,
\bop_1], \bop_2, \dots, \bop_k) \nonumber \\
&+\Sym_k(\bop_1, [\aop, \bop_2], \dots, \bop_k)+ \cdots +
\Sym_k(\bop_1, \bop_2, \dots, [\aop, \bop_k])\ .
\end{align}
\end{prop}
\begin{proof}
The thesis follows from the algebraic identity
\begin{align*}
[\aop,\, \bop_1 \bop_2 \cdots \bop_k]=\ &[\aop, \bop_1] \bop_2
\cdots \bop_k+ \bop_1 [\aop, \bop_2] \bop_3\cdots
\bop_k\nonumber \\
&+\cdots + \bop_1 \bop_2 \cdots \bop_{k-1}[\aop, \bop_k]\ ,
\end{align*}
after symmetrization with respect to indexes $1, \dots,k$ and
rearrangement of the terms of the sum.
\end{proof}

\begin{lem} \label{lem1}
We have
\begin{align}\label{flem0}
\aop\diamond(\bop_1\diamond \bop_2)=\ &\Sym_{3}
(\aop, \bop_1, \bop_2) \nonumber \\
&+\frac 1{12} \big([[\aop, \bop_1],\bop_2]+[[\aop, \bop_2],
\bop_1]\big)
\end{align}
and
\begin{align}
&\aop\diamond\Sym_3(\bop_1, \bop_2, \bop_3)= \Sym_{4}
(\aop, \bop_1, \bop_2, \bop_3) \nonumber \\
&+\frac 1{12} \big\{\bop_1\diamond \big([[\aop, \bop_2], \bop_3] +
[[\aop, \bop_3], \bop_2]\big) +\bop_2\diamond \big([[\aop,
\bop_1], \bop_3] + [[\aop, \bop_3], \bop_1]\big) \nonumber\\
&+\bop_3\diamond \big([[\aop, \bop_1], \bop_2] + [[\aop, \bop_2],
\bop_1]\big)\big\} \,.
\end{align}
The above identities are particular cases of the general formula
\begin{align} \label{flem1}
\aop\diamond\Sym_k(\bop_1, \bop_2, \dots, \bop_k)=\ &\Sym_{k+1}
(\aop, \bop_1, \bop_2, \dots, \bop_k)\nonumber \\
&+ \sum_{h=1}^{[k/2]} c_h C_{k,h} (\aop; \bop_1, \bop_2, \dots,
\bop_k) \,,
\end{align}
where $[k/2]$ indicates the integer part of $k/2$. In the above
formula we put
\begin{align} \label{flem2}
&C_{k,h} (\aop; \bop_1, \bop_2, \dots, \bop_k) \\
:=&\sum_{i_1 \neq i_2 \neq \dots \neq i_{2h}} \Sym_{k-2h+1}
\left([^{2h} \aop, \bop_{i_1}], \bop_{i_2}], \dots],
\bop_{i_{2h}}], \bop_{j_1}, \bop_{j_2}, \dots,
\bop_{j_{k-2h}}\right)\ , \nonumber
\end{align}
where the symbol $[^{2h}$ stands for $2h$ open square brackets
$[[[\cdots [$ in a row, and indexes $j_1, \dots, j_{k-2h}$ have to
be taken so that $(i_1, \dots, i_{2h}) \cup (j_1, \dots, j_{k-2h})
= (1, \dots, k)$. Finally, the coefficients $c_h$ in (\ref{flem1})
are determined recursively by the relations
\begin{equation} \label{coef}
c_1=\frac 1{12}\,, \qquad c_h=-\frac 1{2h+1} \sum_{i=1}^{h-1} c_i
c_{h-i}\quad \mbox{for }h>1\,.
\end{equation}
It follows in particular that
\[
c_2=-\frac 1{720}\,, \quad c_3=\frac 1{30\,240}\,, \quad
c_4=-\frac 1{1\,209\,600}\,, \quad c_5=\frac 1{47\,900\,160}\,,
\quad \dots
\]
\end{lem}
Using (\ref{coef}) it can be shown that $c_h= B_{2h}/(2h)!$,
$h=1,2,\dots$, where $B_h$ are the so-called Bernoulli numbers.
\begin{proof}
For any set of elements $\cop_0, \cop_1, \dots, \cop_k \in
\mathfrak{A}$, let us define
\begin{align} \label{ak}
A_{k} (\cop_0; \cop_1, \cop_2, \dots, \cop_k) :=\ &\cop_0\diamond
\Sym_k(\cop_1, \cop_2, \dots, \cop_k) \nonumber \\
&-\Sym_{k+1} (\cop_0, \cop_1, \cop_2, \dots, \cop_k) \,.
\end{align}
The above expression is obviously symmetric with respect to any
permutations of $\cop_1, \cop_2, \dots, \cop_k$. By symmetry we
have also
\begin{align}\label{s1}
&\cop_0\diamond \Sym_k(\cop_1, \cop_2, \dots, \cop_k)+ \cop
_1\diamond \Sym_k(\cop_0, \cop_2, \dots, \cop_k) \nonumber\\
&+\cop_2\diamond \Sym_k(\cop_1, \cop_0, \cop _3, \dots, \cop_k)
+\cdots +\cop_k\diamond
\Sym_k(\cop_1, \dots, \cop_{k-1}, \cop_0)\nonumber\\
=\ &(k+1) \Sym_{k+1} (\cop_0, \cop_1, \cop_2, \dots, \cop_k) \,,
\end{align}
whence
\begin{align}\label{s2}
&A_{k}(\cop_0;\cop_1, \cop_2, \dots, \cop_k)+ A_{k}(\cop_1;
\cop_0, \cop_2, \dots, \cop_k) \nonumber \\
&+A_{k}(\cop_2; \cop_1, \cop_0,\cop_3, \dots, \cop_k)+ \dots
+A_{k}(\cop_k; \cop_1, \dots, \cop_{k-1}, \cop_0) =0\,.
\end{align}

We shall prove by induction that
\begin{equation} \label{t1}
A_{k} (\aop; \bop_1, \bop_2, \dots, \bop_k)= \sum_{h=1}^{[k/2]}
c_h C_{k,h} (\aop; \bop_1, \bop_2, \dots, \bop_k)\,,
\end{equation}
which is equivalent to (\ref{flem1}). For $k=2$ we have by direct
calculation
\begin{align*}
A_{2} (\aop; \bop_1, \bop_2)=\ &\frac 1{12}(\aop \bop_1 \bop_2 +
\bop_1 \bop_2 \aop + \aop \bop_2 \bop_1 +\bop_2 \bop_1 \aop) \\
&-\frac 16 (\bop_1 \aop \bop_2 + \bop_2 \aop \bop_1) \\
=\ &\frac 1{12}\big([\aop, \bop_1] \bop_2 + \bop_1 [\bop_2, \aop]
+ [\aop, \bop_2] \bop_1 +\bop_2 [\bop_1, \aop]\big) \\
=\ &\frac 1{12} \big([[\aop, \bop_1],\bop_2]+[[\aop, \bop_2],
\bop_1]\big) \,,
\end{align*}
which is equivalent to (\ref{flem0}). This also implies
\begin{align}\label{distr}
&\bop_1\diamond(\bop_2\diamond \aop)- \bop_2\diamond(\bop_1
\diamond\aop) \nonumber \\
=\ &\frac 1{12}\big(2[[\bop_1, \bop_2], \aop] + [[\bop_1, \aop],
\bop_2]- [[\bop_2, \aop], \bop_1]\big) = \frac 14 [[\bop_1,
\bop_2], \aop] \,,
\end{align}
where the last equality follows from Jacobi identity.

According to (\ref{ak}) we can write
\begin{align}\label{flem3}
\aop\diamond \Sym_k(\bop_1, \bop_2, \dots, \bop_k))=\ &\aop
\diamond \big( \bop_1\diamond
\Sym_{k-1}(\bop_2, \dots, \bop_k)\big) \nonumber\\
&-\aop\diamond A_{k-1}(\bop_1; \bop_2, \dots, \bop_k)\,.
\end{align}
From (\ref{distr}), (\ref{ak}) and proposition \ref{pc1}, it
follows that
\begin{align*}
&\aop\diamond \big(\bop_1\diamond \Sym_{k-1}(\bop_2, \dots, \bop_k)\big) \\
=\ &\bop_1\diamond \big(\aop\diamond \Sym_{k-1}(\bop_2, \dots,
\bop_k)\big) +\frac 14 [[\aop, \bop_1], \Sym_{k-1}(\bop_2, \dots,
\bop_k)]\\
=\ &\bop_1\diamond \Sym_{k}(\aop,\bop_2, \dots, \bop_k)+
\bop_1\diamond A_{k-1}(\aop;\bop_2, \dots, \bop_k) \\
&+\frac 14 \big\{\Sym_{k-1}([[\aop, \bop_1], \bop_2], \bop_3,
\dots, \bop_k) +(\Sym_{k-1}(\bop_2, [[\aop, \bop_1], \bop_3],
\dots, \bop_k) \\
&+\cdots+ (\Sym_{k-1}(\bop_2, \dots, \bop_{k-1}, [[\aop, \bop_1],
\bop_k])\big\}\,.
\end{align*}
Let us substitute this expression into the right-hand side of
(\ref{flem3}), and then symmetrize with respect to $\bop_1, \dots,
\bop_k$. Using (\ref{ak}), (\ref{s1}) and (\ref{s2}) we obtain
\begin{align} \label{recur}
&(k+1)A_{k} (\aop; \bop_1, \bop_2, \dots, \bop_k) = \bop_1
\diamond A_{k-1} (\aop; \bop_2, \dots, \bop_k) \nonumber \\
&+\bop_2 \diamond A_{k-1} (\aop; \bop_1, \bop_3, \dots, \bop_k)
+\cdots +\bop_k \diamond A_{k-1} (\aop; \bop_1, \bop_2, \dots,
\bop_{k-1}) \nonumber \\
&+\frac 14 \sum_{i_1\neq i_2} \Sym_{k-1}([[\aop, \bop_{i_1}],
\bop_{i_2}], \bop_{j_1}, \dots, \bop_{j_{k-2}}) \,,
\end{align}
where $(i_1, i_2) \cup (j_1, \dots, j_{k-2}) = (1, \dots, k)$.

Let us now use the hypothesis of induction, and assume that all
elements $A_h$ have the form (\ref{t1}) for $h=2, 3, \dots, k-1$.
We want to begin by showing that $A_{k} (\aop; \bop_1, \bop_2,
\dots, \bop_k)$ contains the term
\begin{equation} \label{cc1} \begin{split}
&c_1 C_{k,1} (\aop; \bop_1, \bop_2, \dots, \bop_k) \\
=\ &\frac 1{12}\sum_{i_1 \neq i_2} \Sym_{k-1} \left([[\aop,
\bop_{i_1}], \bop_{i_2}], \bop_{j_1}, \bop_{j_2}, \dots,
\bop_{j_{k-2}}\right)\,.\end{split}
\end{equation}
For simplicity of notation, let us consider in the above sum only
the term for $i_1=k$, $i_2=k-1$. Using (\ref{ak}) and (\ref{s2})
we have
\begin{align*}
&\bop_1\diamond \Sym_{k-2}([[\aop, \bop_{k}],\bop_{k-1}], \bop_2,
\bop_3, \dots,\bop_{k-2}) \\
&+ \bop_2\diamond \Sym_{k-2}([[\aop, \bop_{k}], \bop_{k-1}],
\bop_1, \bop_3, \dots, \bop_{k-2}) \\
&+\ \cdots+ \bop_{k-2}\diamond \Sym_{k-2}([[\aop, \bop_{k}],
\bop_{k-1}], \bop_1, \bop_2, \dots, \bop_{k-3}) \\
=\ &(k-2)\Sym_{k-1}([[\aop,
\bop_{k}],\bop_{k-1}], \bop_1, \bop_2, \dots, \bop_{k-2})\\
&-A_{k-2}([[\aop, \bop_{k}], \bop_{k-1}];\bop_1, \bop_2, \dots,
\bop_{k-2})\,.
\end{align*}
Therefore, using the induction hypothesis in (\ref{recur}), we
easily see that $A_{k} (\aop;$ $\bop_1, \bop_2, \dots, \bop_k)$
contains the term
\[
c'_1 \Sym_{k-1} ([[\aop, \bop_{k}],\bop_{k-1}], \bop_1, \bop_2,
\dots,\bop_{k-2})\,,
\]
where
\[
c'_1=\frac 1{k+1}\left[(k-2)c_1 +\frac 14 \right] \,.
\]
Using the first of (\ref{coef}) it follows that
\[
c'_1= \frac 1{12}= c_1 \,.
\]
The full expression (\ref{cc1}) can then be obtained by repeating
the above argument for all the other terms of the sum.

Similarly, for $1<h\leq [k/2]$, let us consider all the terms on
the right-hand side of (\ref{recur}) which can contribute to
$A_{k} (\aop; \bop_1, \bop_2, \dots, \bop_k)$ a piece of the form
\begin{equation}\label{alpha}
\alpha\Sym_{k-2h+1} \left([^{2h} \aop, \bop_{k}], \bop_{k-1}],
\dots], \bop_{k-2h+1}], \bop_{1}, \bop_{2}, \dots,
\bop_{k-2h}\right)\,,
\end{equation}
where $\alpha$ is a numerical coefficient. For any $j$ such that
$1\leq j\leq h$, using (\ref{ak}) and (\ref{s2}) we have
\begin{align*}
&\bop_1\diamond \Sym_{k-2j}([^{2j}\aop, \bop_{k}],\bop_{k-1}],
\dots], \bop_{k-2j+1}], \bop_{2}, \bop_{3}, \dots,
\bop_{k-2j}) \\
&+ \bop_2\diamond \Sym_{k-2j}([^{2j}\aop, \bop_{k}],\bop_{k-1}],
\dots], \bop_{k-2j+1}], \bop_{1}, \bop_{3}, \dots,
\bop_{k-2j}) +\cdots \\
&+ \bop_{k-2j}\diamond \Sym_{k-2j}([^{2j}\aop,
\bop_{k}],\bop_{k-1}], \dots], \bop_{k-2j+1}], \bop_{1}, \bop_{2},
\dots, \bop_{k-2j-1}) \\
=\ &(k-2j)\Sym_{k-2j+1}([^{2j}\aop, \bop_{k}],\bop_{k-1}], \dots],
\bop_{k-2j+1}], \bop_{1}, \bop_{2}, \dots,
\bop_{k-2j})\\
&-A_{k-2j}([^{2j}\aop, \bop_{k}],\bop_{k-1}], \dots],
\bop_{k-2j+1}]; \bop_{1}, \bop_{2}, \dots, \bop_{k-2j})\,.
\end{align*}
We see that the above expression contains a term of type
(\ref{alpha}) with
\[
\alpha=\begin{cases} -c_{h-j} &\mbox{if } 1\leq j<h \,,
\\
k-2h &\mbox{if } j=h \,.\end{cases}
\]
Therefore, we find using (\ref{recur}) that $A_{k} (\aop; \bop_1,
\bop_2, \dots, \bop_k)$ contains the term
\[
c'_{h} \Sym_{k-2h+1} \left([^{2h} \aop, \bop_{k}], \bop_{k-1}],
\dots], \bop_{k-2h+1}], \bop_{1}, \bop_{2}, \dots,
\bop_{k-2h}\right) \,,
\]
where
\[
c'_h= \frac 1{k+1}\left[-c_1 c_{h-1} -c_2 c_{h-2}- \cdots -
c_{h-1} c_{1}+ (k-2h)c_h \right] \,.
\]
Using the second of (\ref{coef}) it follows that $c'_h= c_h$, in
agreement with (\ref{t1}). The lemma is thus completely proved.
\end{proof}

\begin{prop}\label{pc2}
We have
\begin{align}\label{c22}
&[\aop_1\diamond\aop_2, \bop_1\diamond\bop_2]= \Sym_3([\aop_1,
\bop_1],\aop_2, \bop_2) + \Sym_3([\aop_1, \bop_2],\aop_2, \bop_1) \nonumber \\
&+\Sym_3([\aop_2, \bop_1],\aop_1, \bop_2) +\Sym_3([\aop_2,
\bop_2],\aop_1, \bop_1) -\frac 1{12}\big([[[\aop_1,\bop_1],
\bop_2],\aop_2]\nonumber \\
&+ [[[\aop_1,\bop_2],\bop_1],\aop_2] +[[[\aop_2,\bop_1], \bop_2],
\aop_1]+ [[[\aop_2,\bop_2], \bop_1],\aop_1]\big)
\end{align}
and
\begin{align}\label{c23}
&[\aop_1\diamond\aop_2, \Sym_3(\bop_1, \bop_2, \bop_3)]= \Sym_4
\left(\aop_1,[\aop_2, \bop_{1}], \bop_{2},\bop_{3}\right)
\nonumber \\
&+\Sym_4 \left(\aop_1,\bop_{1},[\aop_2, \bop_{2}], \bop_{3}\right)
+ \Sym_4\left(\aop_1, \bop_{1}, \bop_{2},[\aop_2,
\bop_{3}]\right)\nonumber \\
&+ \frac 1{12}\sum_{\pi\in\Pi_3}\big( [[\aop_1, \bop_{\pi(1)}],
\bop_{\pi(2)}]\diamond [\aop_2, \bop_{\pi(3)}] -
[[[\aop_1,\bop_{\pi(1)}],\bop_{\pi(2)}],
\aop_2]\diamond \bop_{\pi(3)} \big)\nonumber \\
&+ \aop_1\!\leftrightarrow \!\aop_2\,,
\end{align}
where $\Pi_3$ is the set of all $6$ permutations $\pi$ of 3
objects, and the symbol $\aop_1\!\leftrightarrow \!\aop_2$ means
interchanging $\aop_1$ and $\aop_2$ in all preceding terms on the
right-hand side of the equality. The above identities are
particular cases of the general formula
\begin{align} \label{fprop3}
&[\aop_1\diamond\aop_2, \Sym_k(\bop_1, \bop_2, \dots, \bop_k)]
\nonumber \\
=\ &\sum_{i=1}^k \Sym_{k+1}(\aop_1, \bop_1, \dots,
\bop_{i-1}, [\aop_2,\bop_i], \bop_{i+1}, \dots, \bop_k) \nonumber\\
&-\sum_{h=1}^{[k/2]} c_h D_{k,h} (\aop_1; \aop_2;\bop_1, \bop_2,
\dots, \bop_k) \nonumber \\
&+ \sum_{h=1}^{[(k-1)/2]} c_h E_{k,h} (\aop_1; \aop_2;\bop_1,
\bop_2, \dots, \bop_k) +\aop_1\!\leftrightarrow \!\aop_2\,,
\end{align}
where
\begin{align} \label{fprop1}
&D_{k,h} (\aop_1; \aop_2;\bop_1, \bop_2, \dots, \bop_k) \\
:=&\sum_{i_1 \neq \dots \neq i_{2h}} \Sym_{k-2h+1} \left([^{2h+1}
\aop_1, \bop_{i_1}], \dots], \bop_{i_{2h}}], \aop_2], \bop_{j_1},
\dots, \bop_{j_{k-2h}}\right) \,,\nonumber
\end{align}
\begin{align} \label{fprop2}
&E_{k,h} (\aop_1; \aop_2;\bop_1, \bop_2, \dots, \bop_k) \\
:=&\sum_{i_1 \neq \dots \neq i_{2h+1}} \Sym_{k-2h+1} \left([^{2h}
\aop_1, \bop_{i_1}], \dots], \bop_{i_{2h}}], [\aop_2,
\bop_{i_{2h+1}}], \bop_{j_1}, \dots, \bop_{j_{k-2h-1}}\right)\,,
\nonumber
\end{align}
and the coefficients $c_h$ are determined recursively by relations
(\ref{coef}). Indexes $j_1, \dots, j_{k-2h}$ in (\ref{fprop1})
have to be taken so that $(i_1, \dots, i_{2h}) \cup (j_1, \dots,
j_{k-2h}) = (1, \dots, k)$. An analogous convention is used in
(\ref{fprop2}).
\end{prop}
\begin{proof}
By applying proposition \ref{pc1} twice and lemma \ref{lem1} we
get
\begin{align} \label{pc2f}
&[\aop_1\diamond\aop_2, \Sym_k(\bop_1, \bop_2, \dots, \bop_k)] \nonumber\\
=\ &\aop_1\diamond[\aop_2, \Sym_k(\bop_1, \bop_2, \dots, \bop_k)]+
\aop_2\diamond[\aop_1, \Sym_k(\bop_1, \bop_2, \dots, \bop_k)] \nonumber\\
=\ &\sum_{i=1}^k \aop_1\diamond \Sym_k(\bop_1, \dots, \bop_{i-1},
[\aop_2, \bop_i], \bop_{i+1},\dots, \bop_k) +\aop_1\!
\leftrightarrow \!\aop_2 \nonumber\\
=\ &\sum_{i=1}^k \Sym_{k+1}(\aop_1, \bop_1, \dots, \bop_{i-1},
[\aop_2, \bop_i], \bop_{i+1},\dots, \bop_k) \\
&+\sum_{h=1}^{[k/2]} c_h \sum_{i=1}^k C_{k,h} (\aop_1; \bop_1,
\dots, \bop_{i-1}, [\aop_2, \bop_i], \bop_{i+1},\dots, \bop_k)
+\aop_1 \!\leftrightarrow \!\aop_2 \,. \nonumber
\end{align}
Let us consider all terms which, according to (\ref{flem2}), are
contained in $\sum_{i=1}^k C_{h,k}(\aop_1;$ $\dots, [\aop_2,
\bop_i], \dots)$ for a given value of $h$. Those terms in which
$[\aop_2, \bop_i]$ does not appears inside the iterated
commutators of formula (\ref{flem2}) give rise to the expressions
$E_{k,h}$ in (\ref{fprop3}). The sum of the terms in which
$[\aop_2, \bop_i]$ appears inside the iterated commutators can
instead be considerably simplified by making use of the identity
\begin{equation} \label{iden}\begin{split}
&[^{2h} \aop_1, [\aop_2,\bop_{i_1}]], \bop_{i_2}], \dots],
\bop_{i_{2h}}]+[^{2h} \aop_1, \bop_{i_1}], [\aop_2,\bop_{i_2}]],
\dots], \bop_{i_{2h}}] \\
&+\cdots+ [^{2h} \aop, \bop_{i_1}], \bop_{i_2}], \dots], [\aop_2,
\bop_{i_{2h}}]] \\
=\ &[^{2h+1} \aop_1,\aop_2], \bop_{i_1}], \bop_{i_2}], \dots],
\bop_{i_{2h}}]- [^{2h+1} \aop, \bop_{i_1}], \bop_{i_2}], \dots],
\bop_{i_{2h}}],\aop_2] \,. \end{split}
\end{equation}
This identity is derived by splitting each term on the left-hand
side according to the formula
\begin{align*}
&[^{m} \aop_1, \bop_{i_1}], \bop_{i_2}], \dots],
\bop_{i_{m-1}}],[\aop_2,\bop_{i_{m}}]] \\
=\ &[^{m+1} \aop_1, \bop_{i_1}], \bop_{i_2}], \dots],
\bop_{i_{m-1}}],\aop_2], \bop_{i_{m}}]\\
&-[^{m+1} \aop_1, \bop_{i_1}], \bop_{i_2}], \dots],
\bop_{i_{m-1}}], \bop_{i_{m}}],\aop_2]\,,
\end{align*}
which follows from Jacobi identity. The first term on the
right-hand side of (\ref{iden}) is then cancelled in (\ref{pc2f})
by the corresponding term of $\aop_1 \!\leftrightarrow \!\aop_2$,
whereas the second one contributes to the expressions $D_{k,h}$ in
(\ref{fprop3}).
\end{proof}

\begin{cor}\label{cc2}
Let $(\aop_1,\aop_2)$ and $(\bop_1, \bop_2, \dots, \bop_k)$ be two
sets of elements of $\mathfrak{A}$ such that
\[
[[\aop_i,\bop_j],\bop_{j'}]=0 \quad \forall \,i=1,2\ \mbox{and
}\forall\, j,j'=1,\dots,k, \ j\neq j'\,.
\]
Then
\begin{align} 
&[\aop_1\diamond\aop_2, \Sym_k(\bop_1, \bop_2, \dots, \bop_k)]
\nonumber \\
=\ &\sum_{i=1}^k \big\{\Sym_{k+1}(\aop_1, \bop_1, \dots,
\bop_{i-1}, [\aop_2,\bop_i], \bop_{i+1}, \dots, \bop_k) \nonumber\\
&+\Sym_{k+1}(\aop_2, \bop_1, \dots, \bop_{i-1}, [\aop_1,\bop_i],
\bop_{i+1}, \dots, \bop_k)\big\} \,.
\end{align}
\end{cor}
\begin{proof}
In this case, all terms $D_{h,k}$ and $E_{h,k}$ respectively
defined by formulas (\ref{fprop1}) and (\ref{fprop2}) are zero.
\end{proof}
A typical case, in which the above corollary can be applied, is
when $[\aop_i,\bop_j]$ is a number (more exactly, a number times
the neutral element ${\cal I}$ of the algebra with respect to the
product) $\forall \,i=1,2$ and $\forall\, j'=1,\dots,k$. Under a
similar hypothesis, it is also possible to obtain a formula for
the product of the symmetrized products of two sets containing
arbitrary numbers of elements of the algebra. In order to write
this formula in a sufficiently compact form, let us introduce the
following notations. For all $m\in \mathbb{N}$ we denote with
$N_m:= (1, 2, \ldots, m)$ the set of the first $m$ natural
numbers. For any $h\leq m$ we denote with $P_{h,m}$ the set of the
parts of $N_m$ which contain exactly $h$ elements. For all $I_h\in
P_{h,m}$ we denote with $I_h^C$ the complementary set of $I_h$,
that is $I_h^C:= N_m \setminus I_h$. The set $I_h^C$ obviously
contains exactly $m-h$ elements. For any set $I=(i_1, \ldots,
i_m)\subset \mathbb{N}$ and any $l\in \mathbb{N}$, we also denote
with $l+I$ the set $(l+i_1, \ldots, l+i_m)$.
\begin{prop} \label{wick}
Let $(\lop_1,\ldots,\lop_l)$ and $(\mop_1, \ldots,$ $\mop_m)$ be
two sets of elements of the algebra $\mathfrak{A}$, such that
$[\lop_i,\mop_j] = d_{ij}$, where $d_{ij}$ is a number for all
$i\in N_l$ and all $j\in N_m$. Then
\begin{align} \label{wsymm}
&\Sym_l(\lop_1, \dots, \lop_l) \Sym_m(\mop_1, \dots, \mop_m)
\nonumber \\
=&\sum_{h=0}^{\min(l,m)} \sum_{I\in P_{h,l}}\ \sum_{J\in P_{h,m}}\
\sum_{\pi\in \Pi_h} \frac 1{2^h} d_{i_1
j_{\pi(1)}}\cdots d_{i_h j_{\pi(h)}} \nonumber\\
&\times\ \Sym_{l+m-2h}(\lop_{i'_1}, \dots, \lop_{i'_{l-h}},
\mop_{j'_1}, \dots, \mop_{j'_{m-h}})\,,
\end{align}
where $I= (i_1, \ldots, i_h )$, $J= (j_1, \ldots, j_h )$, $I^C=
(i'_1, \ldots, i'_{l-h})$ and $J^C= (j'_1, \ldots,$ $j'_{m-h})$.
\end{prop}
For instance, for $l=2$ and $m=3$ the above proposition asserts
that
\begin{align*}
&\Sym_2(\lop_1, \lop_2) \Sym_3(\mop_1, \mop_2, \mop_3)
=\Sym_5(\lop_1, \lop_2, \mop_1, \mop_2, \mop_3) \\
&+\frac 12 \big[ d_{11}\Sym_3(\lop_2, \mop_2, \mop_3)
+d_{12}\Sym_3(\lop_2, \mop_1,\mop_3) \\
&+d_{13}\Sym_3(\lop_2, \mop_1, \mop_2)+ d_{21}\Sym_3(\lop_1,
\mop_2, \mop_3)\\
&+d_{22}\Sym_3(\lop_1, \mop_1, \mop_3) +
d_{23}\Sym_3(\lop_1, \mop_1, \mop_2) \big] \\
&+\frac 14 \big[(d_{11} d_{22}+ d_{12} d_{21}) \mop_3 + (d_{11}
d_{23}+ d_{13} d_{21}) \mop_2 + (d_{12} d_{23}+ d_{13} d_{22})
\mop_1 \big] \ .
\end{align*}
Note that no hypothesis has been made on commutators of type
$[\lop_i, \lop_j]$ and $[\mop_i,\mop_j]$.
\begin{proof}
Let us consider the set of elements $(\nop_1,\ldots,\nop_{l+m})$,
where $\nop_i=\lop_i$ for $i=1,\ldots,l$ and $\nop_{l+j}=\mop_j$
for $j=1,\ldots,m$. Let $\tilde\Pi_{l,m}$ be the set of all
permutations $\sigma$ of $l+m$ objects, which do not change the
order of both the group of the first $l$ elements and the group of
the last $m$ elements. More formally, $\tilde\Pi_{l,m}$ is the set
of all $(l+m)!/(l!m!)$ permutations $\sigma\in \Pi_{l+m}$
satisfying the following two conditions:
\renewcommand{\theenumi}{\roman{enumi}}
\begin{enumerate}
\item $\sigma^{-1}(i) < \sigma^{-1}(j)\ \forall\,i,j$ such that
$1\leq i<j \leq l$;

\item $\sigma^{-1}(l+h) < \sigma^{-1}(l+k)\ \forall\, h,k$ such
that $1\leq h<k \leq m$.
\end{enumerate}
\renewcommand{\theenumi}{\arabic{enumi}}
Introducing for $i\in N_l$ and $j\in N_m$ the quantities
\begin{equation} \label{dtilde}
\tilde d^{\sigma}_{ij}= \begin{cases} 0
&\mbox{if }\ \sigma^{-1}(i)<\sigma^{-1}(l+j) \\
d_{ij}&\mbox{if }\ \sigma^{-1}(i)>\sigma^{-1}(l+j) \ ,\end{cases}
\end{equation}
it is easy to prove by induction that, for any $\sigma\in
\tilde\Pi_{m,l}$,
\begin{align} \label{transp}
&\lop_1 \cdots \lop_l\mop_1 \cdots \mop_m =\sum_{h=0}^{\min(l,m)}
\sum_{I\in P_{h,l}}\ \sum_{J\in P_{h,m}}\ \sum_{\pi\in \Pi_h}
\tilde d^{\sigma}_{i_1 j_{\pi(1)}}\cdots \tilde
d^{\sigma}_{i_h j_{\pi(h)}} \nonumber \\
&\times \nop_{\sigma(k_1)} \cdots \nop_{\sigma(k_{l+m-2h})}\ ,
\end{align}
where $I= (i_1, \ldots, i_h )$, $J= (j_1, \ldots, j_h )$,
$(\sigma(k_1), \ldots, \sigma(k_{l+m-2h}))= I^C \cup\ (l+J^C)$ and
$k_1<k_2< \ldots <k_{l+m-2h}$. For instance, for $l=2$, $m=3$ and
$\sigma =(31425)$, the above equality becomes
\begin{align*}
\lop_1 \lop_2\mop_1 \mop_2 \mop_3 =\ &\mop_1 \lop_1 \mop_2 \lop_2
\mop_3 +d_{11}\mop_2 \lop_2 \mop_3 \\
&+d_{21}\lop_1 \mop_2 \mop_3 +d_{22}\mop_1 \lop_1 \mop_3 +d_{11}
d_{22}\mop_3 \ .
\end{align*}

We are obviously allowed to replace the right-hand side of
(\ref{transp}) with its average with respect to all
$(l+m)!/(l!m!)$ distinct permutations $\sigma\in \tilde\Pi_{l,
m}$. If we then symmetrize both members of the equation with
respect to all possible permutations of the $l$ elements $\lop_i$
and the $m$ elements $\mop_j$, we obtain on the right-hand side an
average over all $(l+m)!$ permutations $\sigma \in\Pi_{l+m}$:
\begin{align} 
&\Sym_l(\lop_1, \dots, \lop_l) \Sym_m(\mop_1, \dots, \mop_m)
\nonumber \\
=\ &\frac 1{(l+m)!} \sum_{h=0}^{\min(l,m)} \sum_{I\in P_{h,l}}\
\sum_{J\in P_{h,m}}\ \sum_{\pi\in \Pi_h}\ \sum_{\sigma
\in\Pi_{l+m}} \tilde d^{\sigma}_{i_1 j_{\pi(1)}}
\cdots \tilde d^{\sigma}_{i_h j_{\pi(h)}} \nonumber\\
&\times \nop_{\sigma(k_1)} \cdots \nop_{\sigma(k_{l+m-2h})}\ .
\end{align}
From this, in order to obtain (\ref{wsymm}) we need only observe
that, as a consequence of (\ref{dtilde}),
\begin{align*}
&\frac 1{(l+m)!} \sum_{\sigma \in\Pi_{l+m}} \tilde d^{\sigma}_{i_1
j_{\pi(1)}}\cdots \tilde d^{\sigma}_{i_h j_{\pi(h)}}
\nop_{\sigma(i'_1)} \cdots
\nop_{\sigma(i'_{l+m-2h})} \\
=\ &\frac 1{2^h} d_{i_1 j_{\pi(1)}}\cdots d_{i_h j_{\pi(h)}}
\Sym_{l+m-2h}(\lop_{i'_1}, \dots, \lop_{i'_{l-h}}, \mop_{j'_1},
\dots, \mop_{j'_{m-h}})\,,
\end{align*}
where $I^C= (i'_1, \ldots, i'_{l-h})$ and $J^C= (j'_1, \ldots,
j'_{m-h})$. In fact, $(l+m)!2^{-h}$ is the number of permutations
$\sigma \in\Pi_{l+m}$ such that $\sigma^{-1}(i_r)> \sigma^{-1}
(l+j_{\pi(r)})$ for all $r\in N_h$.
\end{proof}

\begin{cor} \label{wick2}
Under the same hypotheses of the preceding proposition, we have
\begin{align} \label{wsymm2}
&[\Sym_l(\lop_1, \dots, \lop_l), \Sym_m(\mop_1, \dots, \mop_m)]
\nonumber \\
=&\sum_{h\in H}\ \sum_{I\in P_{h,l}}\ \sum_{J\in P_{h,m}}\
\sum_{\pi\in \Pi_h} \frac 1{2^{h-1}} d_{i_1
j_{\pi(1)}}\cdots d_{i_h j_{\pi(h)}} \nonumber\\
&\times\ \Sym_{l+m-2h}(\lop_{i'_1}, \dots, \lop_{i'_{l-h}},
\mop_{j'_1}, \dots, \mop_{j'_{m-h}})\,,
\end{align}
where the notation is the same as that of (\ref{wsymm}), and $H$
is the set of all odd integers $p$ such that $1\leq p\leq
\min(l,m)$.
\end{cor}
\begin{proof}
Since $[\mop_i,\lop_j] = -d_{ji}$, it is easy to see that the
commutator on the left-hand side of (\ref{wsymm2}) is twice the
sum of the terms on the right-hand side of (\ref{wsymm}) which
correspond to odd values of $h$.
\end{proof}

\section{Correspondence by symmetrization between functions on a
Poisson manifold and abstract operators}\label{corresp}


Let us consider the following abstract construction. Let $M$ be a
Poisson manifold. We recall that a Poisson manifold is a manifold
such that an operation called ``Poisson bracket'' is defined in
the space of the functions defined on this manifold. This
operation has the following properties: it is bilinear,
skewsymmetric, it satisfies the identity of Jacobi and the rule of
Leibniz. An important example of Poisson manifold is a symplectic
manifold $M=M^{2N}$ of arbitrary dimension $2N$. Moreover, let
$\aalg$ be an associative algebra of operators with composition as
product.

Let $B=(B_1, \ldots, B_l)$ be a set of linearly independent
functions on the Poisson manifold $M$, which generate a finite
dimensional Lie algebra with respect to Poisson brackets.
Correspondingly, let ${\cal B}=({\cal B}_1, \ldots, {\cal B}_l)$
be a set of linearly independent operators which generate a finite
Lie subalgebra of $\mathfrak{A}$. We suppose that the
correspondence $B_i \to {\cal B}_i$ defines an isomorphism between
Lie algebras.
\begin{defn} \label{symmetr}
Let $P=P(B)$ be a polynomial of $l$ variables. We call the
operator $P^{\rm sym}=P^{\rm sym}(\bop) :=P^{\rm sym}(B)|_{B=
\bop}$, where $P^{\rm sym}$ is the symmetrization of $P$ according
to definition \ref{symm0}, the {\it symmetrization} with respect
to the operators $\bop_1, \ldots, \bop_l$ of the polynomial
$P=P(B_1, \ldots, B_l)$.
\end{defn}

We shall consider initially two different cases, ``constant'' and
``linear''.

\begin{enumerate}
\item In the constant case we have
\begin{equation} \label{lie1}
\{B_i,B_j\}=c_{ij} \,, \quad [\bop_i,\bop_j]=c_{ij}\,, \qquad i,j=
1, \dots, l\,,
\end{equation}
where $c_{ij}=-c_{ji}$ are constant numbers.

\item In the linear case we have
\begin{equation} \label{lie2}
\{B_i,B_j\}= \sum_{k=1}^l c_{ij}^k B_k \,, \quad [\bop_i,\bop_j]=
\sum_{k=1}^l c_{ij}^k \bop_k\,, \qquad i,j= 1, \dots, l\,,
\end{equation}
where $c_{ij}^k=-c_{ji}^k$ are constant numbers. Note that Jacobi
identity implies that
\[
\sum_{k=1}^l \left(c_{ij}^k c_{kh}^m +c_{hi}^k c_{kj}^m +c_{jh}^k
c_{ki}^m \right) =0 \quad\forall\, i,j,h,m=1, \dots, l\,.
\]
\end{enumerate}

If $F=F(B)$ and $H=H(B)$ are polynomials in $B$ of arbitrary
degree with constant coefficients, then $\{H,F\}=G$, where $G$ is
a function on $M$ which too can be represented as a polynomial in
$B$. Such a polynomial representation is unique when $B=(B_1,
\ldots, B_l)$ is a set of polynomially independent functions on
the Poisson manifold $M$. Polynomial independence means that if $P
(B)=0$, where $P$ is a polynomial with constant coefficients, then
$P=0$, i.e., all the coefficients of $P$ are zero. For example, if
the functions $B$ are functionally independent, i.e., their
differentials are linearly independent almost everywhere on $M$,
then they are in particular also polynomially independent.

In any case, the application of linearity and Leibniz rule to
Poisson brackets univocally determines a polynomial representation
for $G=\{H,F\}$. One is therefore free to choose this particular
representation, even when the set $B$ is not polynomially
independent. More precisely, let us consider the map $B: M \to
\rn^l$ such that $B(x) = (B_1(x), \dots, B_l(x))\,\forall \, x\in
M$. This map naturally induces a Poisson structure on $N=\rn^l$.
For any pair of functions $f,g \in C^\infty (\rn^l)$ and
$\forall\, y\in \rn^l$ we have:
\begin{enumerate}
\item In the constant case
\begin{equation} \label{pnc}
\{f, g\}_N (y)= \sum_{i,j=1}^l c_{ij}\frac {\partial f}{\partial
y_i}(y) \frac {\partial g}{\partial y_j}(y) \,; 
\end{equation}
\item In the linear case
\begin{equation} \label{pnl}
\{f, g\}_N (y)= \sum_{i,j,k=1}^l c^k_{ij} y_k\frac {\partial
f}{\partial y_i}(y) \frac {\partial g}{\partial y_j}(y)\,,
\end{equation}
\end{enumerate}
where $\{\,,\}_N$ denote Poisson brackets on $N= \rn^l$. It is
obvious that, in both cases, if $f$ and $g$ are polynomial
functions, then $\{f,g\}_N$ is also a polynomial function. It is
convenient to denote also with $B$ (instead of $y$) the
coordinates on the Poisson manifold $N$.

\begin{defn}\label{leibsym}
Let $F=F(B)$ and $H=H(B)$ be two polynomials in $B$. We say that
the polynomial $G=G(B)= \{H,F\}_N$, obtained with the procedure
described above, is the {\it Leibniz representation with respect
to $B$} of the Poisson bracket $\{H, F\}$. Moreover, we call the
symmetrization $G^{\rm sym}$ of $G$ with respect to $\bop$ the
{\it Leibniz symmetrization with respect to $\bop$} of $\{H, F\}$.
\end{defn}
Obviously we have $G(B)\equiv\{H, F\}$ everywhere in the Poisson
manifold $M$. The name ``Leibniz representation'' is due to the
fact that, in the case of polynomials, the expression of $\{f,
g\}_N$, provided by formulas (\ref{pnc}) or (\ref{pnl}), can also
simply be obtained from (\ref{lie1}) or (\ref{lie2}), by repeated
application of linearity and Leibniz rule to Poisson brackets.

Our goal in this section is to find classes of polynomial
functions $H(B)$ of the functions $B$, such that their Poisson
brackets with any polynomial function $F=F(B)$ is converted by the
operation of Leibniz symmetrization into the commutator of the
corresponding operators:
\begin{equation*} 
[H^{\rm sym},F^{\rm sym}] =\{H,F\}_N^{\rm sym} \ .
\end{equation*}

Let us then consider a function $H$ which has in the two cases
constant and linear considered above the following forms:
\begin{enumerate}
\item In the constant case $H$ is a polynomial of degree two in
the functions $B$:
\begin{equation} \label{h1}
H=\sum_{i\leq j} a_{ij} B_i B_j + \sum_j b_j B_j + c \ .
\end{equation}
\item In the linear case $H$ is a polynomial of degree one, that
is a linear nonhomogeneous function of $B$:
\begin{equation} \label{h2}
H=\sum_j b_j B_j + c \ .
\end{equation}
\end{enumerate}
In the two above equations the coefficients $a_{ij}$, $b_j$ and
$c$ are constants.

\begin{prop} \label{poiss}
Let one of the two following hypotheses be satisfied:
\begin{enumerate}
\item Condition (\ref{lie1}) on Poisson brackets (constant case)
and condition (\ref{h1}) on the degree of $H$;

\item Condition (\ref{lie2}) on Poisson brackets (linear case) and
condition (\ref{h2}) on the degree of $H$.
\end{enumerate}
Suppose also that we have an isomorphism of Lie algebras defined
by the correspondence $B \to {\cal B}$.

Let $F=F(B)$ be an arbitrary polynomial in $B$. Then the
commutator of the operators $H^{\rm sym}$ and $F^{\rm sym}$,
obtained by symmetrization of the polynomials $H$ and $F$
respectively, is equal to the Leibniz symmetrization of the
Poisson bracket of these functions:
\begin{equation} \label{algeb}
[H^{\rm sym},F^{\rm sym}] =\{H,F\}_N^{\rm sym} \ .
\end{equation}
In particular, if the Leibniz representation of $\{H,F\}$ is the
vanishing polynomial, i.e., $\{H,F\}_N=0$, then $[H^{\rm
sym},F^{\rm sym}]= 0$.
\end{prop}
\begin{proof}
It is clearly enough to prove the proposition when the functions
$H$ and $F$ are monomials. In the general situation the thesis
will then follow immediately by linearity. Let us then suppose
that $F=B_{j_1}\cdots B_{j_p}$.

In case 1 let us suppose that $H= B_{h}B_{k}$. Applying corollary
\ref{cc2}, Leibniz rule for Poisson brackets, and relations
(\ref{lie1}), we get
\begin{align*}
&[H^{\rm sym},F^{\rm sym}]= [\bop_{h}\diamond \bop_{k},
\Sym_m(\bop_{j_1}, \dots, \bop_{j_m})] \\
=\ & c_{h j_1} \Sym_m(\bop_{k}, \bop_{j_2}, \bop_{j_3}, \dots,
\bop_{j_m})+ c_{h j_2}\Sym_m(\bop_{k}, \bop_{j_1},
\bop_{j_3}, \dots, \bop_{j_m}) \\
&+ \cdots + c_{h j_m}\Sym_m(\bop_{k}, \bop_{j_1}, \bop_{j_2},
\dots, \bop_{j_{m-1}}) \\
&+ c_{k j_1}\Sym_m(\bop_{h}, \bop_{j_2}, \bop_{j_3}, \dots,
\bop_{j_m})+ c_{k j_2}\Sym_m(\bop_{h}, \bop_{j_1}, \bop_{j_3},
\dots, \bop_{j_m}) \\
&+ \cdots + c_{k j_m}\Sym_m(\bop_{h}, \bop_{j_1}, \bop_{j_2},
\dots, \bop_{j_{m-1}}) \\
=\ & \{B_{h}B_{k},B_{j_1}\cdots B_{j_m}\}_N^{\rm sym}
=\{H,F\}_N^{\rm sym} \ .
\end{align*}

Similarly, in case 2 let us suppose that $H= B_{i}$. Applying
proposition \ref{pc1}, Leibniz rule to Poisson brackets, and
relations (\ref{lie2}), we get
\begin{align*}
&[H^{\rm sym},F^{\rm sym}]= [\bop_{i},\Sym_p(\bop_{j_1}, \dots,
\bop_{j_p})] \nonumber \\
=\ &\sum_k \big[ c_{i j_1 }^{k} \Sym_p(\bop_{k}, \bop_{j_2},
\bop_{j_3}, \dots, \bop_{j_p}) 
+ c_{i j_2 }^{k} \Sym_p (\bop_{j_1},
\bop_k, \bop_{j_3}, \dots, \bop_{j_p}) \nonumber \\
&+ \cdots + c_{i j_p }^{k} \Sym_p(\bop_{j_1}, \bop_{j_2},\dots,
\bop_{j_{p-1}}, \bop_k) \big]\\
=\ &\{B_{i},\,B_{j_1}\cdots
B_{j_p}\}_N^{\rm sym}=\{H,F\}_N^{\rm sym} \,. 
\end{align*}
\end{proof}

\begin{rem}
In case 1, if $H$ has not the form (\ref{h1}), then in general
equality (\ref{algeb}) is no longer valid. Let us in fact identify
the operators $\lop$ and $\mop$ of corollary \ref{wick2} with the
operators $\bop$ of proposition \ref{poiss}, and let us calculate
$[H^{\rm sym},F^{\rm sym}]$ by means of formula (\ref{wsymm2}).
One easily sees that $\{H,F\}_N^{\rm sym}$ corresponds to the term
for $h=1$ on the right-hand side of (\ref{wsymm2}). But, if $H$
contains also terms of degree greater than two in $B$, then there
are in general also terms with $h\geq 3$ which contribute to the
commutator.

Similarly, equality (\ref{algeb}) generally fails in case 2 if $H$
has not the form (\ref{h2}). In fact, if $H$ contains a quadratic
term in $B$, then $[H^{\rm sym},F^{\rm sym}]$ can be calculated by
means of proposition \ref{pc2}. One then sees that $\{H,F\}_N^{\rm
sym}$ corresponds to only the first sum on the right-hand side of
(\ref{fprop3}). In general, however, also the two remaining sums,
containing the terms $D_{h,k}$ and $E_{h,k}$, give nonvanishing
contributions to the commutator.
\end{rem}

\begin{cor} \label{poiss2}
Let $B$ be a set of polynomially independent functions on the
Poisson manifold $M$. Let one of the two following hypotheses be
satisfied:
\begin{enumerate}
\item Condition (\ref{lie1}) on Poisson brackets (constant case)
and condition (\ref{h1}) on the degree of $H$;

\item Condition (\ref{lie2}) on Poisson brackets (linear case) and
condition (\ref{h2}) on the degree of $H$.
\end{enumerate}
Suppose also that we have an isomorphism of Lie algebras defined
by the correspondence $B \to {\cal B}$.

Let $F=F(B)$ be a polynomial in $B$ such that $\{H,F\}=0$
everywhere on $M$. Then
\[
[H^{\rm sym},F^{\rm sym}]= 0\,.
\]
\end{cor}
\begin{proof}
Since $B$ is a polynomially independent set, from the equality
$\{H,F\}=0$ everywhere on $M$ it follows that the Leibniz
representation of $\{H,F\}$ is the vanishing polynomial. The
thesis then follows from proposition \ref{poiss}.
\end{proof}

\begin{rem}\label{moyal}
Let us consider the case in which $B=(x,p)$ is the set of
canonical coordinates in the linear space $\mathbb{R}^{2n}_{xp}$,
and $\bop=(x,\hat p=\partial/\partial x)$ is its canonical
quantization, 
In this case, proposition \ref{poiss} for case 1 (constant) also
directly follows from the useful formula of Moyal brackets
\cite{moyalp}, which can in turn be derived from corollary
\ref{wick2}. We write this formula in the following form:
\[
[H^{\rm sym}, F^{\rm sym}]=G^{\rm sym}\,,
\]
where
\[
G= \sum_{k\in \nn}\ \sum_{|\alpha+ \beta|=2k+1}\frac
{(-1)^{|\beta|}} {2^{2k}\alpha ! \beta !}\,\frac
{\partial^{|\alpha + \beta|} H}{\partial x^\alpha
\partial p^\beta}\, \frac{\partial^{|\alpha + \beta|}F }{\partial
x^\beta \partial p^\alpha} \ .
\]
The sum in the above formula is over all vectors $\alpha
=(\alpha_1, \dots, \alpha_n)\in \zn_+^n$, $\beta =(\beta_1, \dots,
\beta_n)\in \zn_+^n$, such that $|\alpha+ \beta|= \sum_{i=1}^n
(\alpha_i +\beta_i)= 2k+1$. Since $H$ and $F$ are polynomials,
only a finite number of terms of the sum are different from zero.
\end{rem}

We are now going to consider a more general case, in which we get
rid of the hypothesis that the functions of set $B$ form a closed
Lie subalgebra with respect to Poisson brackets. This new case, to
which we shall refer as case 3 (general), can be precisely
formulated in the following way.
\renewcommand{\theenumi}{\roman{enumi}}
\begin{enumerate}
\item \label{gen1} Suppose that there exists a set $T$ of $m$
linearly independent functions, $T= (T_1, \dots, T_m)$, $T_i: M
\to {\mathbb R} \ \forall\ i=1, \dots, m$, and a subset
$B\subseteq T$, $B=(B_1, \dots, B_l)$, with $B_i = T_i$ for $i=1,
\dots, l$, $l \leq m$, such that the Poisson brackets between the
functions of set $B$ are:
\begin{equation} \label{nonlinp}
\{B_i, B_j\}= \sum_{s=1}^m c_{ij}^s T_s \,, \qquad 1\leq i<j\leq l
\ ,
\end{equation}
where $c_{ij}^s= -c_{ji}^s$ are constant coefficients. Note that
we do not suppose that the functions of set $T$ are dependent on
$B$. Moreover, neither the functions of set $B$ nor the functions
of set $T$ are assumed to be closed with respect to Poisson
brackets. However, in practical situations one encounters
sometimes sets of functions $B$ satisfying nonlinear Poisson
bracket relations. In such cases one is able to write formulas of
the form (\ref{nonlinp}), in which the elements of set $T$ are
nonlinear functions of $B$.

\item \label{gen2} Suppose also that there exist a set ${\cal T}$
of $m$ elements of an associative Lie algebra $\mathfrak{A}$ of
operators, and a subset $\bop\subseteq {\cal T}$, $\bop=(\bop_1,
\dots, \bop_l)$, with ${\cal B}_i={\cal T}_i$ for $i=1, \dots, l$,
such that the commutation relations between the operators of set
${\cal B}$ are
\begin{equation} \label{nonlinc}
[{\cal B}_i, {\cal B}_j]= \sum_{s=1}^m c_{ij}^s {\cal T}_s \,,
\qquad 1\leq i<j\leq l \ ,
\end{equation}
where $c_{ij}^s$ are the same constants coefficients as in formula
(\ref{nonlinp}).
\end{enumerate}
\renewcommand{\theenumi}{\arabic{enumi}}

In this case the map $T:M \to \rn^m$ defined by the set $T$ does
not induce a Poisson structure on the set $N=\rn^m$. Nevertheless,
by means of a simple modification of formula (\ref{pnl}), one is
able to define an operation $\{\,,\}_N$, which associates with any
pair of functions $f,g \in C^\infty (\rn^l)$ a function $\{f,g\}_N
\in C^\infty (\rn^m)$. More precisely, $\forall\, y\in N=\rn^m$ we
define
\begin{equation} \label{pnl2}
\{f, g\}_N (y)= \sum_{i,j=1}^l \sum_{k=1}^m c^k_{ij} y_k\frac
{\partial
f}{\partial y_i}(y) \frac {\partial g}{\partial y_j}(y)\,.
\end{equation}
Obviously, if $f$ and $g$ are polynomial functions, then
$\{f,g\}_N$ is also a polynomial function. It may be convenient in
this case to denote with $T$ (instead of $y$) the coordinates on
the Poisson manifold $N=\rn^m$, and with $B$ the first $l$
components of $T$.

\begin{defn}\label{leibsym2}
Let $F=F(B)$ and $H=H(B)$ be two polynomials in $B$. We say that
the polynomial $G=G(T)= \{H,F\}_N$, obtained with the procedure
described above, is the {\it Leibniz representation with respect
to $T$} of the Poisson bracket $\{H, F\}$. Moreover, we call the
symmetrization $G^{\rm sym}$ of $G$ with respect to $\ttop$ the
{\it Leibniz symmetrization with respect to $\ttop$} of $\{H,
F\}$.
\end{defn}
Obviously we have $G(T)\equiv\{H, F\}$ everywhere in the Poisson
manifold $M$. Also in this case, the name ``Leibniz
representation'' is due to the fact that, in the case of
polynomials, the expression of $\{f, g\}_N$, provided by formula
(\ref{pnl2}), can also simply be obtained from (\ref{nonlinp}) by
repeated application of linearity and Leibniz rule to Poisson
brackets.

\begin{prop} \label{notclos}
In the general case 3 specified by hypotheses i and ii given
above, see formulas (\ref{nonlinp})--(\ref{nonlinc}), consider the
function
\begin{equation} \label{h3}
H(B)=\sum_{j=1}^l b_j B_j + c
\end{equation}
and the corresponding operator
\begin{equation} \label{h3op}
{\cal H}=\sum_{j=1}^l b_j {\cal B}_j + c \,.
\end{equation}
The coefficients $b_j$ and $c$ are the same in the two formulas
above. Let $F=F(B)$ be an arbitrary polynomial, and let $F^{\rm
sym}$ denote its symmetrization with respect to $\bop$. Then the
commutator of the operators $\hop$ and $F^{\rm sym}$ is equal to
the Leibniz symmetrization with respect to ${\cal T}$ of the
Poisson bracket $\{H, F\}$:
\begin{equation}\label{algebnl}
[{\cal H},F^{\rm sym}] =\{H,F\}_N^{\rm sym} \ .
\end{equation}
In particular, if the Leibniz representation of $\{H, F\}$ with
respect to $T$ is the vanishing polynomial, i.e., $\{H, F\}_N=0$,
then $[\hop,F^{\rm sym}] =0$.
\end{prop}
\begin{proof}
The proof is similar to that of proposition \ref{poiss} for case
2. For $H=B_i$ and $F=B_{j_1}\cdots B_{j_p}$ we have
\begin{align*}
&[{\cal H},F^{\rm sym}]= [\bop_{i},\Sym_p(\bop_{j_1}, \dots,
\bop_{j_p})] \nonumber \\
=\ &\sum_k \big[ c_{i j_1 }^{k} \Sym_p(\ttop_{k}, \ttop_{j_2},
\ttop_{j_3}, \dots, \ttop_{j_p}) 
+ c_{i j_2 }^{k} \Sym_p (\ttop_{j_1},
\ttop_k, \ttop_{j_3}, \dots, \ttop_{j_p}) \nonumber \\
&+ \cdots + c_{i j_p }^{k} \Sym_p(\ttop_{j_1}, \ttop_{j_2},\dots,
\ttop_{j_{p-1}}, \ttop_k) \big]\\
=\ &\{B_{i},\,B_{j_1}\cdots B_{j_p}\}_N^{\rm sym}=\{H,F\}_N^{\rm
sym} \,.
\end{align*}
The general result follows by linearity.
\end{proof}

The following corollary is the analogue of corollary \ref{poiss2}.
\begin{cor} \label{notclos2}
Let hypotheses i--ii above be satisfied, see formulas
(\ref{nonlinp})--(\ref{nonlinc}). Consider the function $H$ and
the operator $\hop$ given by formulas (\ref{h3}) and (\ref{h3op})
respectively. In addition, suppose that $T$ is a set of
polynomially independent functions on the Poisson manifold $M$.
Let $F=F(B)$ be a polynomial in $B$ such that $\{H,F\}=0$, and let
$F^{\rm sym}$ denote its symmetrization with respect to $\bop$.
Then
\begin{equation}
[{\cal H},F^{\rm sym}] =0\ .
\end{equation}
\end{cor}

\begin{rem} \label{poissclos}
Very often the set $T$ is locally closed with respect to Poisson
brackets, that is {\it locally Poisson-closed}. (If this set is
not closed, usually it makes sense to consider its Poisson
closure, i.e., its extension by means of finite iterations of
Poisson brackets.) Let us consider for simplicity a set $T$ which
is globally Poisson-closed, i.e., there exist globally uniquely
defined functions $c_{ij}$ of $m$ variables such that $\{T_i,
T_j\} = c_{ij}(T)$ at all points of $M$, $1\leq i < j \leq m$. In
this case, using the map $T:M \to \rn^m$, defined by the set $T=
(T_1, \dots, T_m)$, one can transfer the Poisson structure of $M$
on its image $T(M)\subseteq \rn^m$. With respect to this Poisson
structure induced on $N:=T(M)$, the map $T$ becomes a Poisson map.
We recall that a map $T: M \to N$, where $M$ and $N$ are Poisson
manifolds, is called a Poisson map if
\[
\{f,g\}_N \circ T= \{f \circ T, g \circ T\}_M
\]
for any pair of functions $f, g \in C^\infty (N)$. Here
$\{\,,\}_M$ and $\{\,,\}_N$ denote Poisson brackets in the two
Poisson manifolds $M$ and $N$ respectively, and the symbol $\circ$
is used to indicate the composition of maps. When $T$ is a
globally Poisson-closed set of functions on $M$, the above formula
univocally determines a Poisson structure on $N$. The Poisson
bracket of two functions $f$ and $g$ on $N$ takes the form
\[
\{f,g\}_N (y) := \sum_{i,j=1}^m c_{ij}(y)\frac {\partial
f}{\partial y_i}(y) \frac {\partial g}{\partial y_j}(y)
\]
for all $y=(y_1, \dots, y_m)\in N\subseteq \rn^m$.
\end{rem}
An example of Poisson-closed set of functions on a Poisson
manifold $M$ is the set of functions $B=(B_1, \dots, B_l)$ in the
linear case 2. The set $B$ is a basis of the $l$-dimensional Lie
algebra $\mathfrak{g}$ defined by the linear Poisson bracket
relations (\ref{lie2}). In this case, the set $B$ defines the
Poisson map $B:M \to N$, where the Poisson manifold $N$ is
isomorphic to the $l$-dimensional linear space $\rn^l$. The
manifold $N$ can be identified with the Lie co-algebra
$\mathfrak{g}^*$, that is with the space which is the linear
conjugate to the Lie algebra $\mathfrak{g}$: $N=\mathfrak{g}^*$.
The Poisson structure on $N$ transferred from $M$ can be written
in intrinsic terms: for any two functions $g,f \in C^\infty
(\mathfrak{g}^*)$ and any point $\xi \in \mathfrak{g}^*$, one
defines $\{g, f\}_\xi := ( [dg|_\xi, df|_\xi], \xi )$, where the
differentials $dg|_\xi$ and $df|_\xi$ are elements of the Lie
algebra $\mathfrak{g}$, and $[\,\,,\,]$ is the commutator in this
Lie algebra. This means that any Lie co-algebra has a natural
structure of Poisson manifold. It has to be noted that the
elements of set $B$, considered as functions on the Poisson
manifold $\galg^*$, are always functionally independent, even when
they are not functionally independent as functions on the Poisson
manifold $M$.
\begin{defn} \label{defcas2}
A {\it Casimir function} on the Lie co-algebra $\galg^*$ is an
invariant of the co-adjoint action of the local Lie group $G$
associated with the Lie algebra $\galg$. It means that $f:\galg^*
\to\rn$ is a Casimir function if $(df|_\xi, \tau_{b, \xi})= 0\
\forall\, \xi \in \galg^*$ and $\forall\ b \in \galg$, where
$\tau_b \in \galg^*$ is defined by the formula $(a,\tau_{b, \xi})=
([b,a], \xi)\ \forall\,a\in \galg$. Hence, for any other function
$g:\galg^* \to\rn$ we can write $\{g, f\}_\xi= ([dg|_\xi,
df|_\xi], \xi) =(df|_\xi, \tau_{b, \xi}) =0$, with $b= dg|_\xi$.
It follows that Casimir functions are in involution with any other
function defined on the Lie co-algebra.
\end{defn}

\begin{rem}
It is known that any function $H:M \to \rn$ on a Poisson manifold
$M$ defines univocally a vector field $X_H$ on $M$, which is
tangent to the symplectic leaves of this manifold $M$. (In local
coordinates $y$ on $M$, this vector field $X_H$ is defined by the
system of differential equations $\dot y= \{H, y\}$.) This vector
field $X_H$ is called {\it hamiltonian with hamiltonian function
$H$}. As in the case of symplectic manifolds, one has $[X_H, X_F]=
X_{\{H,F\}}$ for any two functions $H$ and $F$ on any Poisson
manifold $M$, where the left-hand side represents the Lie bracket
of vector fields. So one has a natural homomorphism of the Lie
algebra of functions on the Poisson manifold $M$ to the Lie
algebra of vector fields on $M$. In particular the structure of a
finite dimensional Lie subalgebra of functions on $M$ transfers to
the vectors fields corresponding to these functions.

Let us consider again a finite-dimensional Lie co-algebra
$\galg^*$, with the intrinsic structure of Poisson manifold
described above. Any linear function $H:\galg^* \to \rn$ can be
considered as an element $a$ of the Lie algebra $\mathfrak{g}$ and
can be written in the form $H=H_a$. In this case, the phase flow
of the hamiltonian vector field $X_H$, $H=H_a$, coincides with a
local one-parametric Lie subgroup of the Lie group $G$ with
co-adjoint action on the Lie co-algebra $\mathfrak{g}^*$: $G\times
\mathfrak{g}^* \to \mathfrak{g}^*$. Here $G$ is the local Lie
group corresponding to the Lie algebra $\mathfrak{g}$, and this
subgroup of $G$ corresponds to the element $a \in \mathfrak{g}$.
If $a \neq 0$, this means that the straight line tangent to this
subgroup at the neutral element $e \in G$ contains the vector $a
\in \mathfrak{g}$.
\end{rem}

Propositions \ref{poiss} and \ref{notclos} are particular cases of
the following more general proposition. Let us suppose that we
have, in addition to the sets $T$ and ${\cal T}$, another set of
functions $D=(D_1, \ldots, D_s)$ on the Poisson manifold $M$, and
another set of operators ${\cal D}=({\cal D}_1, \ldots, {\cal
D}_s)$ of the associative algebra $\mathfrak{A}$, such that
$\{D_i,D_j\}= \{D_i,T_k\}=0$, $[{\cal D}_i,{\cal D}_j]=[{\cal
D}_i,{\cal T}_k]=0$ for $i,j=1,\ldots, s$ and $k=1,\ldots,m$. In
cases 1 and 2 one can take $T=B$, ${\cal T}= {\cal B}$ and $m= l$.
In case 1 let us also suppose that the coefficients $a_{ij}$,
$b_j$ and $c$ of formula (\ref{h1}), and the coefficients $c_{ij}$
of formula (\ref{lie1}), are polynomial functions of $D$.
Similarly, in cases 2 and 3, let us suppose that the coefficients
$b_j$ and $c$ of formulas (\ref{h2}) and (\ref{h3}), and the
coefficients $c_{ij}^{k}$ of formulas (\ref{lie2}) and
(\ref{nonlinp}), are polynomial functions of $D$.

For this situation we now generalize the definition of
symmetrization. Consider the linear space $P[D,T]$ of all
polynomials $P=P(D,T)$ of $s$ variables $D$ and $m$ variables $T$.
Consider the linear map sym: $P[D,T] \to \mathfrak{A}$, which
associates with an arbitrary monomial $Q=D_1^{\alpha_1} \cdots
D_s^{\alpha_s}T_{i_1} \cdots T_{i_k}$ the operator
\[
Q^{\rm sym}:= {\cal D}_1^{\alpha_1} \cdots {\cal D}_s^{\alpha_s}
\Sym_k(\ttop_{i_1}, \dots, \ttop_{i_k})\ .
\]
Here the $\alpha_i$ are integers numbers, $\alpha_i\geq 0$. The
monomials $Q$ form a basis in the linear space $P[D,T]$. Hence the
mapping $Q\mapsto Q^{\rm sym}$ defines a linear map sym: $P\mapsto
P^{\rm sym}$ on the full space $P[D,T]$.

\begin{defn}\label{dsym}
We call the operator $P^{\rm sym}= P^{\rm sym}(\dop, \ttop) \in
\mathfrak{A}$, associated with the polynomial $P(D,T)$ by the
linear map defined above, the {\it symmetrization} with respect to
the operators ${\cal T}_1, \ldots, {\cal T}_l$ of the polynomial
$P=P(D,T)$.
\end{defn}

We observe that the functions of $D$ can practically be treated as
constant in all the algebraic manipulations that were necessary to
prove propositions \ref{poiss} and \ref{notclos}. Therefore:
\begin{prop}\label{nonconst}
In this more general situation, propositions \ref{poiss} and
\ref{notclos} remain valid. The symmetrization of all terms of
formulas (\ref{algeb}) and (\ref{algebnl}) is now performed with
respect to the operators of the set ${\cal T}$.
\end{prop}

\section{Application to quantization}\label{apq}

The term ``quantization'' is here used for convenience. We will
discuss the construction, from functions defined on the Poisson
manifold $M$, of differential operators acting on functions
defined on the manifold $K$. These operators are not necessarily
connected with quantum mechanics, for example we can consider the
linear operator with nonconstant coefficients which defines the
heat equation 
We suppose that in this
construction the Poisson brackets of functions are converted into
Lie brackets of operators which are the ``quantization'' of these
functions. As a rule, the manifold $M$ is the cotangent bundle of
$K$, $M= T^*K$, but this is not necessary. This transformation of
functions into operators is based on the assumption that there
exists a correspondence between some finite-dimensional algebras
of functions on $M$ and of operators on $K$. This correspondence
under some assumptions is extended to products of functions which
correspond to the composition of operators.

So, let us consider a particularly interesting case for us, in
which $\bop_1, \ldots, \bop_l$ are linear differential operators
in the variables $x=(x_1, \ldots, x_n)$. More precisely, using the
notation of 
\cite{part1}, we suppose that $\bop_i\in {\cal O}={\cal O}_K$,
$i=1, \ldots, l$, where $K$ is a domain of $\mathbb{R}^n_x$. Let
these operators form a Lie algebra with respect to commutators. As
above, we also suppose the existence of a set $B=(B_1, \ldots,
B_l)$ of functions on the Poisson manifold $M$, which forms a
basis of a Lie algebra with respect to Poisson brackets. We
suppose that there exists an isomorphism ${\cal I}$ of Lie
algebras, which is defined by the correspondence $Q:B_i \to
\bop_i$, $i=1,\ldots,l$, between their bases. Let us consider an
arbitrary polynomial $P$ of $l$ variables, and let $P^{\rm
sym}(\bop)$ be the symmetrization of $P$ with respect to the
operators $\bop$, see definition \ref{symmetr}.

\begin{defn} \label{symq}
In this case, the correspondence $Q$ and the isomorphism ${\cal
I}$ defined by $Q$ is called {\it basic quantization}. The
operator $P^{\rm sym}(\bop)$ is called {\it quantization by
symmetrization} of the polynomial $P(B)$ of variables $B=(B_1,
\ldots,B_l)$, or {\it symmetric quantization of $P(B)$ with
respect to the set $B$}.
\end{defn}

Let us consider the important example $M^{2n}=
\mathbb{R}^{2n}_{xp}$, $B=(x,p)$ and $\bop=(x,\hat p)$, where
$\hat p=\partial/\partial x$. We have
\begin{equation} \label{cpb}
\{x_i,x_j\}= 0\,, \qquad \{p_i,p_j\}= 0\,, \qquad \{p_i,x_j\}=
\delta_{ij}
\end{equation}
and
\begin{equation} \label{ccr}
[x_i,x_j]= 0\,, \qquad [\hat p_i,\hat p_j]= 0\,, \qquad [\hat p_i,
x_j] = \delta_{ij}
\end{equation}
for $i,j= 1, \dots, n$, where $\delta_{ij}$ is the Kr\"onecker
symbol. Hence the sets $(x,p)$ and $(x, \hat p)$ are bases in two
$2n$-dimensional Lie algebras, and the correspondence between
these bases defines an isomorphism of Lie algebras.
\begin{defn}\label{can}
In this case the basic quantization, given by the correspondence
$Q_{\rm c}:(x,p)\to (x,\hat p)$ and the isomorphism ${\cal I}_{\rm
c}={\cal I}(Q_{\rm c})$, is called {\it canonical quantization}.
\end{defn}
Note that canonical quantization corresponds to the standard
quantization of the set $(x,p)$ according to
the definition given in \cite{part1}. However, the standard
quantization of a polynomial $P(x,p)$ in general does not coincide
with its symmetric quantization $P^{\rm sym}(x,\partial/\partial
x)$ constructed according to definition \ref{symq}.

Let us suppose that $B=(B_1,\ldots,B_l)$ is a set of arbitrary
functions on some Poisson manifold $M$. Consider the set of
operators $\fop=(\fop_1,\ldots, \fop_r;$ $\fop_{r+1},\ldots,
\fop_s)$ on $\mathbb{R}^{n}_{x}$, which are obtained from a given
basic quantization $Q:B\to\bop$, $B=(B_1,\ldots,B_l)$, by
symmetrization (with respect to $B$) of the set of polynomials
$P=(P_1, \ldots,P_r;$ $P_{r+1}, \ldots, P_s)$, $P_i=P_i(B)$. We
can thus write $\fop_i = P_i^{\rm sym} (\bop)$. We suppose that
\begin{equation}\label{pb}
\{P_i(B),P_j(B)\}_N=0\,,\qquad i=1, \ldots,r,\ j=1,\ldots, s\,,
\end{equation}
where $\{\,,\}_N$ denote the Leibniz representation with respect
to $B$ of Poisson brackets, see definition \ref{leibsym}. Consider
the case 1 (constant): $\{B_i,B_j\}=c_{ij}$.

\begin{prop} \label{appl1}
Let us suppose additionally that the polynomials of the set $P$
satisfy at least one of the two following \medskip conditions:
\renewcommand{\theenumi}{\alph{enumi}}
\begin{enumerate}
\item $\deg P_i(B)\leq 2$, $i=1,\ldots, r$.
\item $\deg P_j(B)\leq 2$, $j=2,\ldots, s$.
\end{enumerate}
\renewcommand{\theenumi}{\arabic{enumi}}
Here and in next proposition $\deg$ means degree with respect to
$B$.

Then the following two statements are true:
\begin{enumerate}
\item $[\fop_i, \fop_j]=0$, $i=1, \ldots,r$, $j=1,\ldots, s$.

\item Let us additionally suppose that we can extract two sets of
polynomials $(P'_1,\dots, P'_{r'})$ $\subseteq (P_1, \dots,
P_{r})$ and $(P'_{r'+1}, \dots, P'_{s'}) \subseteq (P_{r+1},
\dots, P_{r+s})$, where $r'+s' =2n$, such that the operators of
the set $\fop'=$ $(\fop'_1, \dots, \fop'_{r'};$ $\fop'_{r'+1},
\dots, \fop'_{s'})$ are quasi-independent, see definition
of quasi-independence in \cite{part1}.

Then the set of operators $\fop'$ is quasi-integrable with $r'$
central integrals $(\fop'_1, \dots, \fop'_{r'})$.
\end{enumerate}
\end{prop}

Let us now consider the case 2 (linear): $\{B_i,B_j\}= \sum_k
c_{ij}^{k}B_k$.

\begin{prop} \label{appl2}
Under the same hypotheses stated before proposition \ref{appl1},
let us suppose that the polynomials of the set $P$ satisfy at
least one of the two following conditions: 
\renewcommand{\theenumi}{\alph{enumi}}
\begin{enumerate}

\item $\deg P_i(B)\leq 1$, $i=1,\ldots, r$.

\item $\deg P_j(B)\leq 1$, $j=2,\ldots, s$.



\end{enumerate}
\renewcommand{\theenumi}{\arabic{enumi}}

Then the same statements 1 and 2 of proposition \ref{appl1} are
true.
\end{prop}

Let us finally consider case 3 (general), see hypotheses
\ref{gen1}--\ref{gen2} before definition \ref{leibsym2}. In this
case we have two sets of functions $B$ and $T$, such that $B
\subseteq T$, and two sets of operators $\bop$ and ${\cal T}$,
such that $\bop \subseteq {\cal T}$. These functions and operators
satisfy the relations (\ref{nonlinc})--(\ref{nonlinp}).
Furthermore, we have a set of polynomials $P=(P_1, \ldots,P_r;$
$P_{r+1}, \ldots, P_s)$, $P_i=P_i(B)$, which satisfy relations
(\ref{pb}), where $\{\,,\}_N$ now denote the Leibniz
representation with respect to $T$ of Poisson brackets, see
definition \ref{leibsym2}. Finally, we have a set of operators
$\fop$ which are obtained from $P$ by symmetrization: $\fop_i =
P_i^{\rm sym} (\bop)$, on the basis of the correspondence $Q: B
\to \bop$. The following proposition is a generalization of
proposition \ref{appl2} in two different respects. One does not
require that the functions of either the set $B$ or $T$ satisfy
linear Poisson brackets relations, and one does not require that
such sets be closed with respect to Poisson brackets.

\begin{prop} \label{appl3}
Under the above hypotheses, let us additionally suppose that the
polynomials of the set $P$ satisfy at least one of the two
conditions a or b of proposition \ref{appl2}. Then the same
statements 1 and 2 of proposition \ref{appl1} are true.
\end{prop}

Propositions \ref{appl1}--\ref{appl2} easily follow from
proposition \ref{poiss}, while the last proposition \ref{appl3}
follows from proposition \ref{notclos}.
\begin{rem}\label{polyi}
Let us suppose that the functions $B_1, \dots, B_l$ are
polynomially independent almost everywhere in $M$. In this case,
the equality $\{P_i(B),P_j(B)\}=0$ in $M$ implies that the Leibniz
representation of $\{P_i(B),P_j(B)\}$ is the vanishing polynomial,
i.e., $\{P_i(B),P_j(B)\}_N=0$. Hence, in propositions
\ref{appl1}--\ref{appl2} one can replace the hypothesis
$\{P_i(B),P_j(B)\}_N=0$ with $\{P_i(B),P_j(B)\}=0$. The same
statement is true for proposition \ref{appl3}, if the functions
$T_1, \dots, T_m$ are polynomially independent almost everywhere
in $M$.
\end{rem}

Let us consider the case of linear relations $\{B_i, B_j\}=
\sum_{s=1}^l c_{ij}^s B_s$, see formula (\ref{lie2}). Let $\galg$
be the Lie algebra formed by linear combinations of the functions
of set $B$. We know already (see section \ref{corresp}) that the
co-algebra $\galg^*$, i.e., the linear space conjugated to
$\galg$, has a natural structure of Poisson manifold. As above,
let us suppose that there is a set of linear operators $\bop =
(\bop_1, \dots, \bop_l)$ such that $[\bop_i, \bop_j]= \sum_{s=1}^l
c_{ij}^s \bop_s$ with the same constants $c_{ij}^s$. Let the
polynomial $C=C(B)$, i.e., $C: \galg^* \to \rn$, be some invariant
of the co-adjoint representation on $\galg^*$ of the local group
$G$ which corresponds to the Lie algebra $\galg$.

\begin{cor} \label{casim2}
In this case the operator $C^{\rm sym}$ commutes with all
operators of set $\bop$: $[C^{\rm sym}, \bop]=0$.
\end{cor}
\begin{proof}
Taking into account that any invariant of the co-adjoint
representation of a group is a Casimir function on $\galg^*$, we
have $\{C, B_i\}_N= 0$ for all functions $B_i$ of set $B$, or
shortly $\{C, B\}_N= 0$, where $\{\,,\}_N$ denotes Poisson bracket
on the co-algebra $\galg^*$. The thesis then follows from
proposition \ref{appl2}, case b.
\end{proof}
Obviously, corollary \ref{casim2} also implies that $[C^{\rm sym},
P(\bop)] =0$ for any noncommutative polynomial $P\in S^{0,l}_N$.

\begin{rem}
According to 
the results of \cite{part1}, if a set $(\fop_1, \dots, \fop_k;$
$\fop_{k+1},$ $\dots, \fop_{2n-k})$ of operators is
quasi-integrable, then the set $(MF_1, \dots,$ $MF_k; MF_{k+1},
\dots,$ $MF_{2n-k})$ of the main parts of their symbols $F_i:=
\fop_i^{\rm smb}$ is also integrable in the usual classical sense.
Therefore, in the situation of statement 2 of propositions
\ref{appl1}--\ref{appl3} ($r'+s'=2n$), the set $MF'$ of the main
parts of the symbols of the operators ${\fop_i'}^{\rm smb}$, $i=1,
\dots, s'$, must be integrable in the classical sense.

Note that classical integrability of a set $(F_1, \dots, F_k;
F_{k+1}, \dots, F_{2n-k})$ implies that this set is (locally)
closed with respect to Poisson brackets. This means that locally,
in some neighborhood of almost every point, one has $\{F_i, F_j\}
= f_{ij}(F)$, where $f_{ij}$ are some functions of $2n-k$
variables and $i,j= 1, \dots, 2n-k$ (see remark \ref{poissclos}).
In particular, in the situation considered above, the set of
functions $MF'$ is integrable and hence Poisson-closed.
\end{rem}

\begin{cor} \label{applcor}
Let us suppose that the hypotheses of one of propositions
\ref{appl1}--\ref{appl3} are true, with $r+s=2n$, and that the
operators $\fop$ are quasi-independent. Let ${\cal H}$ be an
operator of class ${\cal O}$ and let the set $({\cal H}, {\cal
F}_{1}, \ldots, {\cal F}_{k})$ be globally dependent. For example,
${\cal H}=S({\cal F}_{1}, \ldots, {\cal F}_{k})$, where $S$ is an
arbitrary polynomial of $k$ variables, that is $S\in {\cal
S}_C^{k}$. We suppose also that $[{\cal H},{\cal F}_{i}]=0$ for
each $i=1,\ldots,2n-k$. In this case, ${\cal H}$ is a
quasi-integrable operator with $k$ central operators.
\end{cor}

\begin{rem} The consideration of general case 3 allows one to go
outside the frame of finite Lie algebras of functions on
symplectic manifolds, and correspondingly of linear differential
operators. Operators forming a basis of a finite Lie algebra
usually are quadratic with respect to $(x, \partial/\partial x)$,
or linear with respect to $\partial/\partial x$. The consideration
of nonlinear commutation relations gives in principle the
possibility of dealing with operators of more general type. An
example of nonlinear commutation relations, arising from the
quantization of a classical system of nonlinear resonant
oscillators, will be considered in 
a following paper.
\end{rem}

\begin{rem} \label{sinx}
The class of systems to which the last corollary can be applied is
too small for many physical applications. For example, one has
often to deal with functions which are not polynomials. Let us
consider the situation of canonical quantization, where $B$ is the
set of canonical coordinates $(p,x)$ (constant case). In this case
it is often necessary to consider functions of $x$ which are not
polynomials. Sometimes one also considers functions of $p$ such as
$\exp (ip)$, and also functions of the form $f(x_1,p_2)$. Of
course, if one considers operators of class ${\cal O}$, that is
polynomials with respect to $p$, then the only non polynomial
functions one can meet are of the form $f(x)$. In this case it is
necessary to use an additional property of the operators under
consideration. Namely, suppose there exist some subsets $\tilde
{\cal B}^i\subseteq {\cal B}$ such that $[\tilde {\cal B}^i,
\tilde {\cal B}^i]=0$ for each $i=1,\dots, l$. This means that all
operators of the set $\tilde {\cal B}^i$ are pairwise commuting.
Suppose also that for functions $f$ of some class (which contains
also non polynomials functions) the operator $f(\tilde {\cal
B}^i)$ is well defined. Then, $[f(\tilde {\cal B}^i),g(\tilde
{\cal B}^i)]=0$ for any functions $f,g$ of this class, and if
$[{\cal B}_j , \tilde {\cal B}^i]=0$ for some ${\cal B}_j \in
{\cal B}$, we have $[{\cal B}_j , f(\tilde {\cal B}^i)]=0$.
\end{rem}

Let us suppose that there exist also a set of functions $D=(D_1,
\ldots, D_s)$ and a set of operators ${\cal D}=({\cal D}_1,
\ldots, {\cal D}_s)$ with the properties considered in proposition
\ref{nonconst}.

\begin{cor}
In this case, statement 1 of propositions \ref{appl1}--\ref{appl3}
remains true when the constants $c_{ij}$ in formula (\ref{lie1})
or $c_{ij}^{k}$ in formulas (\ref{lie2}) or (\ref{nonlinp}) are
replaced by functions $c_{ij}(D)$ or $c_{ij}^{k}(D)$ respectively,
and the coefficients of polynomials $P_i(B)$ are functions of $D$.
The symmetrization of these polynomials is performed with respect
to operators ${\cal T}$ according to definition \ref{dsym}.
\end{cor}
The proof is an easy application of proposition \ref{nonconst}.

\end{document}